\documentclass[aps,pre,preprint,showpacs,amsmath,amssymb,amsfonts,preprintnumbers]{revtex4}
\input amssym.def

\def\Cb{{\Bbb C}}

\begin{document}
\baselineskip18pt
\thispagestyle{empty}

\begin{center}\bf   Entanglement of Multipartite Schmidt-correlated States \end{center}
\vskip 1mm

\begin{center}
Ming-Jing Zhao$^{1}$, Shao-Ming Fei$^{1,2}$ and  Zhi-Xi Wang$^{1}$

\vspace{2ex}

\begin{minipage}{5in}

\small $~^{1}$ {\small Department of Mathematics, Capital Normal University, Beijing 100037}

%{\small $~^{2}$ Institut f\"ur Angewandte Mathematik, Universit\"at Bonn, D-53115}

{\small $~^{2}$ Max-Planck-Institute for Mathematics in the Sciences, 04103 Leipzig}

\end{minipage}
\end{center}

\vskip 2mm
\parbox{14cm}
{\footnotesize\quad We generalize the
Schmidt-correlated states to multipartite systems. The related
equivalence under SLOCC, the separability, entanglement witness,
entanglement measures of negativity, concurrence and relative entropy
are investigated in detail for the generalized Schmidt-correlated states.}

\bigskip
{{\it Keywords}: {\footnotesize  Schmidt-correlated state;
separability; entanglement measure}}

\section{Introduction}

Quantum entanglement is of special importance in quantum information
processing and responsible for many quantum tasks such as
teleportation, dense coding, key distribution, error correction etc.
\cite{M.A.Nielsen}. Entanglement of bipartite states have been
extensively studied. There have been many necessary or/and
sufficient conditions on separability for certain states such as
PPT, reduction, majorization, realignment etc. \cite{Peres A.,M.
Horodecki,Nielsen,Rudolph,K.Chen}. There are also some well defined
measures of entanglement for bipartite states, e.g. entanglement of
formation (EoF) \cite{BDSW,Horo-Bruss-Plenioreviews}, concurrence
\cite{con}, negativity \cite {G.Vidal} and relative entropy \cite{V.
Vedral}. For two-qubit case EoF is a monotonically increasing
function of concurrence and an elegant formula of concurrence was
derived analytically by Wootters in \cite{Wootters98}. For higher
dimensional case, due to the extremizations involved in the
calculation, only a few of explicit analytic formulae for EoF and
concurrence have been found for some special symmetric states
\cite{th-fjlw-fwzh,Rungta03}, and attention has been payed to
possible lower bounds of the EoF and concurrence e.g.
\cite{Chen-Albeverio-Fei,Chen-Albeverio-Fei1,breuer}.

Among the quantum states, the Schmidt-correlated (SC) states are of
special properties. They are the mixtures of pure states, sharing
the same Schmidt bases. It was first appeared in \cite{Rains}, named
as maximally correlated state. For any classical measurement related
to the SC states, two observers will always obtain the same result.
Two SC states can always be optimally discriminated locally. The
maximally entangled states (Bell state) can always be expressed in
Schmidt correlated form. SC states naturally appear in a bipartite
system dynamics with additive integrals of motion \cite{Khasin06}.
Hence, these states form an important class of mixed states from a
quantum dynamical perspective. Bipartite SC states have been studied
in \cite{V. Vedral,E.M.Rains,S. Virmani,Yi,M.Khasin}. The Relative
Entropy of SC bipartite states has been investigated in \cite{V.
Vedral,E.M.Rains,Yi}. The properties of negativity for SC states are
studied in \cite{M.Khasin}. In this paper we generalize the
Schmidt-correlated bipartite states to multipartite ones. We study
the separability and entanglement of the multipartite SC states.

\section{Definition}

 We first give the definition of multipartite SC state in arbitrary
dimension. We call a $k$-partite state $\rho $ in $ C^N \otimes C^N
\otimes\cdots \otimes C^N$ an SC state if it can be expressed as
\begin{equation}\label{ksc}
\rho = \sum_{m,n=0 }^{N-1} a_{mn} |m \cdots m \rangle \langle n \cdots n|,
\end{equation}
where $\sum_{m=0}^{N-1} a_{mm}=1$. (\ref{ksc}) is a direct
generalization of bipartite ones. If $A_i$ $(1\leq{i} \leq{k})$ is
the observer associated with the $i$-th sub-system, then $A_1,
\cdots, A_k $ will always obtain the same result for any classical
measurement.

It has been shown in \cite{Yi} that for a general mixed state $\rho
= \displaystyle {\sum_{m,n=0}^{N-1}a_{mn}|m \rangle \langle n|}$,
there exists an ensemble of pure states $\{p_i,|\Phi_i\rangle\}$
realizing $\rho$, where $ |\Phi_i \rangle $ is of the form $ |\Phi_i
\rangle = \sum_m \sqrt{a_{mm}} e^{i \Theta_m^{(i)}} |m \rangle $.
This result can be easily generalized to multipartite case: the SC
state (\ref{ksc}) could be realized by an ensemble $\{p_i,| \Phi_i
\rangle\}$, where $\displaystyle {| \Phi_i \rangle = \sum_m \sqrt
{a_{mm}} e^{ i \Theta_m^{(i)}} |m \cdots m \rangle} $, with $a_{mm}$
given in (\ref{ksc}). Furthermore, $\rho$ can only be realized by
such ensembles. In fact, if there exists an ensemble
$\{p_i,|\Phi_i\rangle\}$ realizing $\rho$ such that $| \Phi_i\rangle
= \sum_{m_1, \cdots, m_k} c_{m_1, \cdots, m_k}^i |m_1, \cdots, m_k
\rangle$ for some $i$ and different $m_1, \cdots, m_k$,
there must be an item $|m_1, \cdots, m_k \rangle \langle m_1,
\cdots, m_k|$ in the presentation of $\rho$, so that $\rho$
is no longer of the form (\ref{ksc}).

Let ${\rm GHZ}(k,N)$ denote the $k$-partite maximally entangled
state
$$ {\rm GHZ}(k,N) = \frac{1}{\sqrt{N}} (|0 \cdots 0 \rangle + |1 \cdots 1\rangle +
\cdots + |N-1, \cdots, N-1 \rangle ).$$ Then $ |\Phi_i \rangle $ is
equivalent to either a fully separable state or ${\rm GHZ}(k,t)~ (0
< t \leq N)$ under stochastic local operation and classical
communication (SLOCC) \cite{FENG,L. Lamata}. Indeed, if $a_{mm}~ (0
\leq m \leq N-1)$ are all zero except one, then $ |\Phi_i \rangle $
is fully separable. If not, let us suppose $a_{00}, \cdots,
a_{(t-1),(t-1)}$ $(1< t \leq N)$ are nonzero and the rest are zero.
Let
\begin{eqnarray*} F&=&\left(
\begin{array}{cccc}
\frac{1}{( \sqrt{t} \sqrt{a_{00}} e^{i\Theta^{(i)}_0 })^{\frac{1}{k}} }& 0 & \cdots & 0  \\
0 & \frac{1}{( \sqrt{t} \sqrt{a_{11}} e^{i\Theta^{(i)}_1 })^{\frac{1}{k}} } & \cdots &  0  \\
\cdots & \cdots & \cdots & \cdots  \\
0 & 0 & \cdots & \frac{1}{( \sqrt{t} \sqrt{a_{t-1,t-1}}
e^{i\Theta^{(i)}_{t-1} })^{\frac{1}{k}} }
\end{array}
\right)\\[3mm]
&=& \frac{1}{( \sqrt{t} \sqrt{a_{00}} e^{i\Theta^{(i)}_0
})^{\frac{1}{k}} } |0 \rangle \langle 0| + \frac{1}{( \sqrt{t}
\sqrt{a_{11}}
e^{i\Theta^{(i)}_1 })^{\frac{1}{k}} } |1 \rangle \langle 1| + \cdots\\[3mm]
&&+ \frac{1}{( \sqrt{t} \sqrt{a_{t-1,t-1}} e^{i\Theta^{(i)}_{t-1}
})^{\frac{1}{k}} } |t-1 \rangle \langle t-1|.
\end{eqnarray*}
Obviously $F$ is invertible. We can get ${\rm GHZ}(k,t) = F \otimes
F \otimes\cdots \otimes F\,| \Phi_i \rangle$.

In \cite{Arun} it has been proven that if the partial inner
product of a basis of any one of the subsystems with the state of a
composite system gives a disentangled basis, then Schmidt
decomposition for a tripartite system exists. This criterion can be
generalized to multipartite states. We can see that $|\Phi _i
\rangle$ are states which have Schmidt decompositions in a
$k$-partite composite system.

\section{Separability of Schmidt-correlated states}

The state (\ref{ksc}) is fully separable if $\rho=\sum_i p_i
\rho_1^{(i)} \otimes \rho_2^{(i)} \otimes\cdots
\otimes\rho_k^{(i)}$, where $\sum_i p_i =1$, $\rho_1^{(i)},
\rho_2^{(i)}, \cdots, \rho_k^{(i)}$ are states in individual
subsystems.

$\textbf{Proposition 1.}$  \ \  State (\ref{ksc}) is fully separable
if and only if it is positive under partial transpositions with
respect to some subsystems.

$\textbf{Proof.}$ Without loss of generality, we choose $\rho ^{T_1}$ as
an example of partial transposition. The other cases are similar
to $\rho^{T_1}$. The proof can be done along the same lines as calculation of
negativity in \cite{M.Khasin}. In fact the eigenvalues of $\rho$ under
partial transpositions with respect to any subsystems are the same to $\rho ^{T_1}$. Hence $\rho$ is
positive under some partial transpositions if and only if $\rho$ is
positive under any partial transpositions. By carrying out some elementary
transformations, $\rho^{T_1} = \sum_{ m,n=0
}^{N-1} a_{mn} |n m \cdots m \rangle \langle m n \cdots n|$, can
be transformed into another matrix $(\rho ^{T_1})^ \prime$,
$$ (\rho ^{T_1} ) ^\prime = \displaystyle \left(
\begin{array}{ccc}
A & 0 &  0  \\
0 & B & 0 \\
0 & 0 & C \\
\end{array}
\right),
$$
where $A={\rm diag} (a_{00},a_{11},...,a_{N-1,N-1})$,
\begin{eqnarray*}
B &=& \displaystyle \left(
\begin{array}{cccccccccc}
0 & a_{10} & 0 & 0 &  \cdots &  0 & 0 & \cdots & 0 & 0\\
a_{01} & 0 & 0& 0 & \cdots & 0 & 0 & \cdots & 0 & 0\\
0 & 0 & 0 & a_{20} & \cdots & 0 & 0 &\cdots &  0 & 0\\
0 & 0 & a_{02} & 0 & \cdots & 0 & 0 &\cdots&  0 & 0\\
\cdots & \cdots & \cdots & \cdots & \cdots & \cdots & \cdots &\cdots & \cdots  & \cdots \\
0 & 0 & 0 & 0 & \cdots & 0 & a_{N-1,0} & \cdots & 0 & 0 \\
0 & 0 & 0 & 0 & \cdots & a_{0,N-1} & 0 & \cdots & 0 & 0 \\
\cdots & \cdots & \cdots & \cdots & \cdots & \cdots & \cdots &\cdots & \cdots  & \cdots \\
0 & 0 & 0 & 0 & \cdots & 0 & 0 & \cdots & 0 & a_{N-1,N-2} \\
0 & 0 & 0 & 0 & \cdots & 0 & 0 & \cdots & a_{N-2,N-1} & 0
\end{array}
\right),
\end{eqnarray*}
and $C$ is a zero matrix.

The nonzero eigenvalues of $ (\rho^{T_1}) ^\prime $ are $
\lambda_0=a_{00}$, $\lambda_1=a_{11}$, $\cdots$,
$\lambda_{N-1}=a_{N-1,N-1}$, $\lambda_{mn}^{\pm}=\pm |a_{mn}|$, $m <
n$, $m, n=0, \cdots, N-1 $, which are also the eigenvalues of $
 \rho^{T_1}$. Therefore $\rho^{T_1} \geq 0$ if and only if $\lambda_{mn}^{\pm} = 0$.
In this case, the only nonzero eigenvalues of $\rho^{T_1}$ are $
\lambda_0, \cdots, \lambda_{N-1} $ with respect to the eigenvectors
$|0\cdots 0 \rangle$, $\cdots$, $|N-1,\cdots, N-1 \rangle$. And
$\rho=\rho^{T_1}=a_{00}|0 \cdots 0 \rangle \langle 0 \cdots 0| +
\cdots + a_{N-1,N-1}|N-1, \cdots, N-1 \rangle \langle N-1, \cdots,
N-1|$. That is, $\rho$ is fully separable.

Conversely, if $\rho$ is fully separable, then $\rho$ is also
bipartite separable in the partition $ 1|2 \cdots k $, $\rho = \sum
_i q_i \rho _1^{(i)} \otimes \rho _{2 \cdots k}^{(i)} $. By Peres'
criterion \cite{Peres A.}, we have $\rho^{T_1} \geq 0 $. $\Box$

Therefore for an SC state $\rho$, $\rho$ is fully separable if and
only if it is positive with respect to some partial transposition,
and $\rho$ is genuinely entangled if and only if it is not positive
with respect to some partial transposition.

Entanglement witness is an operator that is designed for
distinguishing between separable and entangled states \cite{M.
Horodecki,B. M. Terhal,Philipp}. A Hermitian operator $W$ is called
an entanglement witness if it has a positive expectation value with
respect to any separable state $\sigma$, $Tr[W \sigma]\geq 0$, while
there exists at least one entangled state $\rho$ such that $Tr[W
\rho]<0$. Next we construct entanglement witness for SC states.

$\textbf{Corollary 1.}$ \ \ Let $| \Psi_{mn} \rangle$ be
eigenvectors associated with the eigenvalues $\lambda_{mn}^-$, $ m <
n$, $m, n=0, \cdots, N-1 $. Then $W= \sum _{m < n} (| \Psi_{mn}
\rangle \langle \Psi_{mn}|)^{T_1}$ is an entanglement witness of
$\rho$.

$\textbf {Proof.}$
For any separable state $\sigma$, we have
$$Tr[W
\sigma] = \sum _{m < n} Tr [(| \Psi_{mn} \rangle \langle
\Psi_{mn}|)^{T_1} \sigma ]=\sum _{m < n} Tr [| \Psi_{mn} \rangle
\langle \Psi_{mn}| \sigma^{T_1}] \geq 0, $$
as $\sigma^{T_1} \geq 0$. For the entangled SC state $\rho$, we get
$$
Tr[W \rho ] = \sum _{m < n} Tr [(| \Psi_{mn} \rangle
\langle \Psi_{mn}|)^{T_1} \rho] = \sum _{m < n} Tr[| \Psi_{mn}
\rangle \langle \Psi_{mn}| \rho^{T_1}] =
 \sum _{m < n} \lambda_{mn}^-<0.
$$
Therefore $W$ is a witness for $\rho$. $\Box$

Any bipartite state $\rho$ in $V_1 \otimes   V_2 $ can be written in
Bloch representation \cite{Julio}. Let $\lambda _i~(0 \leq i \leq
N^2-2)$ be the generators of $SU(N)$, $0 \leq j < k \leq N-1$,
$$
\lambda _{i}= \sqrt {\frac{2}{(i+1)(i+2)}}( \sum _{a=0}^i|a \rangle
\langle a|-(i+1)|i+1 \rangle \langle i+1|),~~~i = 0,\cdots,  N-2,
$$
$$\lambda _i=|j \rangle \langle k|+|k \rangle \langle j|,~~~i=N-1, \cdots, \frac{(N+2)(N-1)}{2}-1,$$
$$ \lambda _i=-i(|j \rangle \langle k| - |k \rangle \langle j|),~~~i =\frac{(N+2)(N-1)}{2}, \cdots,  N^2-2. $$
Any  bipartite state $\rho$ can be written as
$$\rho =
\frac{1}{MN}(I_M \otimes I_N + r_i \lambda _i \otimes  I_N + s_j I_M
\otimes  \lambda _j + t_{ij} \lambda _i \otimes
 \lambda _j ),$$
 where  $ M=\dim V_1$, $ N=\dim V_2$,
 $r_i=\frac {M}{2} Tr( \rho \lambda _i\otimes   I_N )$,
$s_j = \frac{N}{2}Tr(\rho I_M \otimes \lambda _j)$,
 $t_{ij} = \frac{MN}{4}Tr(\rho \lambda _i \otimes   \lambda _j)$. A pure bipartite state
 is separable if and only if
 $t_{ij}=r_i s_j$ for any $i$ and $j$ \cite{Julio}.

For a multipartite SC state $\rho$, we express it in Bloch
representation as a bipartite decomposition. Assuming that a given
$k$-partite SC state is divided in to two subsystems, the first one
consisting of $l$ parties and the second one of $(k-l)$ parties.
Thus the subsystems are defined on $N^l$-dimensional and
$N^{k-l}$-dimensional Hilbert spaces, respectively. This means in
turn that to describe such a state in the Bloch representation one
can take generators of $SU(N^l)$ and $SU(N^{k-l})$ as bases in
respective subsystems. By calculation we find $r_i=0$ for $i \geq
N^l-1$, $s_i=0$ for $i \geq N^{k-l}-1$, $t_{ij}=0$ for $i \geq
N^l-1$, $0 \leq j \leq N^{k-l}-2$, or $0 \leq i \leq N^{l}-2$, $j
\geq N^{k-l}-1$. According to our Proposition 1, (\ref{ksc}) is
separable if and only if $\rho = \sum_{m=0 }^{N-1} a_{mm} |m \cdots
m \rangle \langle m \cdots m|$. We have $t_{ij}=0$ for a separable
SC state, where $i \geq N^l-1$, $j \geq N^{k-l}-1$. Conversely, if
$t_{ij}=0$ for all $i \geq N^l-1$, $j \geq N^{k-l}-1$, then
$a_{mn}=0$ for $m\neq n$. Hence $\rho$ is fully separable.
Consequently, we get

$\textbf{Corollary 2.}$ \ \ A multipartite SC state, expressed in
Bloch representation as a bipartite decomposition, is fully
separable if and only if $t_{ij}=0$ holds for $i \geq N^l-1$, $j
\geq N^{k-l}-1$.

Another complementary operational separability criterion is called
realignment criterion \cite{K.Chen,Rudolph} which detects even
bound entangled states. It says that for any separable state
the realigned matrix $R(\rho)$ of $\rho$ satisfies $\parallel R(\rho)
\parallel \leq 1$, where $\| \rho \| \equiv tr
\sqrt{\rho^ \dag \rho}$, $R(\rho)$ is given by $R(\rho) _{ij,
kl}=\rho_{ik, jl}$ with $i$ and $j$ the row and column
indices for the first subsystem and $k$ and $l$ the
indices for the second subsystem. For multipartite SC states, let us first view them as
bipartite states in $\Cb^N \otimes \Cb^M$.

 $\textbf{Proposition 2.}$ \ \ For any SC state $\rho$ in $\it
V_1 \otimes   V_2 \otimes  \cdots \otimes   V_k$, $\dim V_i = N~ (1
\leq i \leq k) $, we have $ 1 \leq \parallel R(\rho) \parallel \leq N $.
A state $\rho$ is fully separable (resp. maximally entangled) if and only if $ \parallel R(\rho) \parallel  =1 $
(resp.  $ \parallel R(\rho) \parallel =N$).

$\textbf{Proof}$.\ \ For multipartite SC states (\ref{ksc}), we have that
$R(\rho)$ is a diagonal matrix with diagonal entries $a_{mn}$ and
some zeros. Thus
\begin{eqnarray*} \|R(\rho)\|= \sum _{m, n=0} ^{N-1} |a_{mn}| \leq
\sum _{m, n=0} ^{N-1} \sqrt{a_{mm}a_{nn}} \leq \sum _{m, n=0} ^{N-1}
\frac{1}{2} (a_{mm}+a_{nn}) =N.
\end{eqnarray*}
The first inequality follows from the positivity of the density
operator and the second one from the inequality of geometric and
arithmetic means.

Here $\parallel R(\rho) \parallel = 1$ implies $a_{mn}=0$ for $m\neq
n$, $m,n=0, \cdots, N-1$. Therefore in this case $\rho$ is fully
separable. When $\parallel R(\rho) \parallel = N$, we obtain
$a_{mn}=\frac{1}{N}$ for all $m, n=0, 1, \cdots, N-1$, i.e. $\rho$
is ${\rm GHZ}(k,N)$. Hence $\rho$ is maximally entangled in any
bipartite decompositions. $\Box$

\section{Entanglement of Schmidt-correlated states}

We now calculate some measures of entanglement for multipartite SC
states.

\subsection{Negativity}

One entanglement measure for bipartite states defined in
\cite{G.Vidal} is the negativity, $N(\rho)= \frac {\|
\rho^{T_1}\|-1}{2}$, which corresponds to the absolute value of the
sum of negative eigenvalues of $\rho^{T_1}$. Now let's have a look
at the negativity of SC states in a bipartite decomposition.

For any SC state $\rho$ in $\it V_1 \otimes   V_2 \otimes  \cdots
\otimes   V_k$, $\dim V_i = N$ $(1 \leq i \leq k) $, it is
straightforward to verify that $N(\rho)=\frac{1}{2} \sum _{m \neq n}
|a_{mn}|=\frac{\|R(\rho) \|-1}{2}$. From Proposition 2 we have $ 0
\leq N(\rho) \leq \frac {N-1} {2}$ associated with any partial
transpositions. If $ N(\rho)=0 $ for some partial transpositions,
then $\rho$ is fully separable. If $ N(\rho)= \frac {N-1}{2}$ with
respect to some partial transpositions, then $\rho $ is the maximally
entangled pure GHZ state.

Indeed, for any $|\Psi \rangle = \sum_ \alpha c_\alpha e_\alpha
^\prime \otimes  e_\alpha ^{\prime \prime }$, $N( \rho) = \frac
{1}{2} [( \sum_ \alpha c_\alpha )^2 -1]$ \cite{G.Vidal}. As an
example for the maximally entangled state ${\rm GHZ}(2,N) = \frac
{1}{\sqrt{N}} ( |00 \rangle +|11 \rangle + \cdots + |N-1,N-1
\rangle)$, we have $N({\rm GHZ}(2,N)) = \frac {N-1}{2}$.

Since SC states are either genuinely entangled or fully separable,
we know that if $ N(\rho) \neq 0 $, then $\rho$ must be genuinely
entangled. But it may be not maximally entangled. For instance, for
states in $\Cb^2 \otimes  \Cb^2 \otimes  \Cb^2 $: $ \rho = \frac{2}{3}[
\frac{1}{2} ( |000 \rangle + |111 \rangle ) (\langle 000| + \langle
111|)] + \frac{1}{3} |000 \rangle \langle 000 | $ and ${\rm
GHZ}(3,2) = \frac{1}{2} ( |000 \rangle + |111 \rangle ) (\langle
000| + \langle 111|)$, we have $ N(\rho) = 1/3$ and
 $N({\rm GHZ}(3,2)) =1/2$ respectively. $\rho$ is genuinely entangled but not maximally entangled.

\subsection{Concurrence}

The concurrence for a bipartite pure state $|\Psi \rangle$ is
defined by $C(|\Psi \rangle)= \sqrt{2(1-Tr \rho_1^2)}$, where the
reduced density matrix $\rho_1$ is given by $\rho_1=Tr_2(|\Psi
\rangle \langle \Psi|)$. The concurrence is then extended to mixed
states $\rho$ by the convex roof, $C(\rho) \equiv \min _{\{p_i,
|\Psi _i \rangle \}} \sum_i p_i C(|\Psi _i \rangle)$, for all
possible ensemble realizations $\rho= \sum _i p_i |\Psi_i \rangle
\langle \Psi_i|$, where $p_i \geq 0$ and $\sum_i p_i=1$. A state
$\rho$ is separable if and only if $C(\rho)=0$.

For a multipartite SC state $\rho$,
\begin{equation}\label{rho}
\rho = \sum_{ m,n=0 }^{N-1} a_{mn} |m \cdots m \rangle \langle n
\cdots n| = \sum _i p_i |\Psi_i \rangle \langle \Psi_i|,
\end{equation}
where, as discussed in section II, $|\Psi _i\rangle$ takes the form
$|\Psi _i \rangle = \sum _m c_m^{(i)} |m \cdots m \rangle$,
$\sum _m |c_m^{(i)}|^2=1$, $a_{mn}=\sum_i p_i c_m^{(i)} c_n^{(i)*}$.
We get that the concurrences of $|\Psi _i \rangle$ are the same for
all reduced density matrices in bipartite decompositions. Let us choose $\rho _1=Tr_{2 \cdots
k}(|\Psi_i \rangle \langle \Psi_i|)$ as an example. We have $\rho
_1= \sum_m |c_m^{(i)}|^2 |m \rangle \langle m|$,
\begin{eqnarray*}
\\
&C(|\Psi _i \rangle)&=\sqrt{2(1- \sum _m |c_m^{(i)}|^4)},
\\
&C(\rho)&=\min _{\{p_i, |\Psi _i \rangle \}} \sum_i p_i
\sqrt{2(1-\sum _m |c_m^{(i)}|^4 )} =2\min _{\{p_i, |\Psi _i \rangle
\}} \sum_i p_i \sqrt{\sum_{m<n} |c_m^{(i)} c_n^{(i)}|^2}.
\end{eqnarray*}

Taking into account that $\sum_m |c_m^{(i)}|^2=1$,
we have $0 \leq C(\rho)\leq \sqrt{2(1-\frac{1}{N})}$.
For the state ${\rm GHZ}(k,N)$, one has
$$
C({\rm GHZ}(k,N))=\sqrt{2(1- \sum _{m=0}^{N-1}\frac{1}{N^2})}=\sqrt{2(1- \frac{1}{N})}.
$$

In \cite{Chen-Albeverio-Fei} an analytical lower bound
for the concurrence of bipartite quantum states in arbitrary
dimension has been derived, which is exact for some mixed quantum states.
For $N$-dimensional SC states $\rho$, the result is reduced to be
$C(\rho) \geq \frac{2\sqrt{2}}{\sqrt{N(N-1)}} N(\rho)$.
In particular, for multi-qubit ($N=2$) SC states
$\rho=\sum_i p_i |\Phi_i \rangle \langle\Phi_i|$, if
$ |\Phi_i \rangle = \alpha _i|0 \cdots 0
\rangle + \beta_i e^{i\theta}|1 \cdots 1 \rangle$, where
$\alpha_i,\beta_i \geq 0$ and $0\leq \theta \leq 2\pi$,
then $C(\rho)=2\min_{\{p_i, |\Psi _i \rangle \}} \sum_i p_i\alpha_i\beta_i
=2N(\rho)$. Namely if all the pure states $|\Phi_i \rangle$ in the
decomposition share the same relative phase $e^{i\theta}$, then the
equality holds, $C(\rho)=2N(\rho)=\|R(\rho)\|-1$.
Moreover, for pure SC qubit states,
we have $C(\rho)=2N(\rho)=\|R(\rho)\|-1$.
For example, for $|\Psi \rangle =\sqrt{\frac{1}{3}}|0 \cdots 0 \rangle +\sqrt{\frac{2}{3}}|1 \cdots 1
\rangle$, we have $C(|\Psi \rangle)=2N(|\Psi \rangle)=\|R(|\Psi \rangle)\|-1
=\frac{2\sqrt{2}}{3}$.

Concerning the concurrence for multipartite states, instead of
bipartite decompositions, there is also a generalized version:
$$
C(|\Psi\rangle)= \sqrt{k-Tr \rho_1^2 -Tr \rho_2^2-\cdots -Tr \rho_k^2},
$$
where the reduced density matrix $\rho_i$ is given by
$\rho_i=Tr_{1,\cdots,i-1,i+1,\cdots,k}(|\Psi \rangle \langle
\Psi|)$, $i=1,2,\cdots,k$. Extending to mixed states $\rho$,
$$
C(\rho) \equiv \min _{\{p_i,
|\Psi _i \rangle \}} \sum_i p_i C(|\Psi _i \rangle),
$$
for all possible ensemble realizations $\rho= \sum _i p_i |\Psi_i \rangle
\langle \Psi_i|$. Similarly we can get $0\leq C(\rho) \leq
\sqrt{k(1-\frac{1}{N})}$ from Lagrange multipliers method. Applying
to the state ${\rm GHZ}(k,N)$, we have
$$
C({\rm GHZ}(k,N))=\sqrt{k(1-\frac{1}{N})}.
$$

\subsection{Relative Entropy}

 Relative Entropy is a well defined measure for multipartite states. For a
density matrix $\rho$,  $ E(\rho) =\min _{ \sigma \in D} S( \rho
\parallel \sigma ) =\min_{ \sigma \in D} Tr [ \rho \log \rho - \rho \log \sigma ]$,
where $D$ is the set of all fully separable states \cite{V. Vedral}.
However, it is not easy to find the
optimal separable state $\sigma ^*$ such that
$E(\rho)=\min_{ \sigma \in D}S( \rho \parallel \sigma ) = S( \rho
\parallel \sigma ^* )$. It has been proven that
$ \sigma ^* =\sum _m a_{mm} | mm \rangle \langle mm |$ is PPT optimal for
bipartite state $\rho = \sum _{mn} a_{mn} | mm \rangle \langle nn | $ \cite{E.M.Rains},
that is, $S(\rho \parallel \sigma ^* )=\min_{ \sigma ^\prime \in P}S( \rho
\parallel \sigma ^\prime ) $, where $P$ is the set of all PPT states.

For a $k$-partite SC state $\rho =
\sum _{mn} a_{mn} | m \cdots m  \rangle \langle n \cdots n |$,
$E(\rho) = min _{ \sigma \in D} S( \rho \parallel \sigma ) = S( \rho
\parallel \sigma ^* )$, where $ \sigma ^* =
\sum _m a_{mm} | m \cdots m \rangle \langle m \cdots m |$, and $D$ is
the set of all fully separable states.
We consider a $k$-partite SC state as a bipartite SC state, for example,
the first subsystem and the rest subsystems,
$ \rho = \sum _{mn} a_{mn} (|m \rangle \otimes |m \cdots m
\rangle) (\langle n| \otimes \langle n \cdots n |)$. Then according to
\cite{E.M.Rains}, $ S(\rho
\parallel \sigma ^* )=\min _{ \sigma ^\prime \in P } S(
\rho\parallel \sigma ^\prime ) $, where $P$ is the set of all PPT states.
Since fully separable states are all positive under partial transpose
with respect to any subsystems, we have $\min _{ \sigma \in D} S( \rho
\parallel \sigma ) \geq \min _{ \sigma  ^\prime
\in P } S( \rho \parallel \sigma ^\prime ) $. Because $ \sigma ^* $
is separable, we get the formula $ E(\rho) = \min _{ \sigma \in D} S( \rho
\parallel \sigma ) = S( \rho \parallel \sigma ^* )$ also for multipartite
SC states.

For the state ${\rm GHZ}(k,N)$, the optimal fully separable state is
just
$$\sigma ^*=\frac{1}{N} \sum_{i=0}^{N-1}|ii \cdots i \rangle
\langle ii \cdots i|.$$
And $E({\rm GHZ}(k,N))=S({\rm GHZ}(k,N) \|
\sigma ^*)= Tr[\rho \log \rho - \rho \log \sigma ^*]=\log N$.

By calculating three kinds of entanglement measures for SC states,
we can see that the entanglement of multipartite maximally entangled
states is independent on the number of subsystems $k$. The
entanglement is only related to the dimensions of the subsystems.

\section{Conclusions}

We have investigated a special kind of multipartite states named SC
states. The sufficient and necessary conditions of separability for
these states have been studied. We have also calculated the
negativity, concurrence and relative entropy of SC states. By
calculating the $N(\rho)$, $C(\rho)$ or $E(\rho)$, the entanglements
of any two SC states can be compared. Moreover, like bipartite SC
states that naturally appear in dynamics with additive integrals of
motion \cite{Khasin06}, the multipartite SC states, which have
always bipartite decompositions, would also form an important class
of mixed states from a quantum dynamical perspective.


\begin{thebibliography}{18}

\bibitem{M.A.Nielsen} M.A. Nielsen and I.L. Chuang, {\it Quantum Computation and Quantum
Information}
(Cambridge Univesity Press, Cambridge, 2000).

\bibitem{Peres A.} A. Peres, Phys. Rev. Lett.  {\bf 77},  1413(1996).

\bibitem{M. Horodecki} M. Horodecki, P. Horodecki and R. Horodecki, Phys. Lett. A  {\bf 223}, 1(1996).

\bibitem{Nielsen}
M. A. Nielsen and J. Kempe, Phys. Rev. Lett. {\bf 86}, 5184(2001).

\bibitem{Rudolph}
O. Rudolph, Quantum Inf. Proc. \textbf{4}, 219(2005).

\bibitem{K.Chen}
K. Chen and L.A. Wu. Quantum Inf. Comput. \textbf{3}, 193(2003).

\bibitem{BDSW} C.H. Bennett, D.P. DiVincenzo, J.A. Smolin, and W.K.
Wootters, Phys. Rev. A \textbf{54}, 3824(1996).

\bibitem{Horo-Bruss-Plenioreviews} M. Horodecki, Quantum Inf. Comp. \textbf{1}, 3(2001);\\
D. Bru\ss , J. Math. Phys. \textbf{43}, 4237(2002);\\
M.B. Plenio and S. Virmani, Quantum Inf. Comp. \textbf{7}, 1(2007).

\bibitem{con}
A. Uhlmann, Phys. Rev. A \textbf{62}, 032307(2000);\\
P. Rungta, V. Buzek, C.M. Caves, M. Hillery and G.J. Milburn, Phys. Rev. A
\textbf{64}, 042315(2001);\\
S. Albeverio and S. M. Fei, J. Opt. B: Quantum Semiclass. Opt. \textbf{3}, 223(2001).

\bibitem{G.Vidal} G. Vidal and R. F. Werner, Phys. Rev. A \textbf{65}, 032314(2002).

\bibitem{V. Vedral} V. Vedral and M.B. Pleniio, Phys Rev A \textbf{57}, 1619(1998).

\bibitem{Wootters98} W.K. Wootters, Phys. Rev. Lett. \textbf{80}, 2245(1998).

\bibitem{th-fjlw-fwzh}
B.M. Terhal, K. Gerd and K.G.H. Vollbrecht, Phys. Rev. Lett.
{\bf 85}, 2625(2000).\\
S.M. Fei, J. Jost, X.Q. Li-Jost and G.F. Wang, Phys. Lett. A {\bf
310}, 333(2003).\\
S.M. Fei, Z.X. Wang and H. Zhao, Phys. Lett. A {\bf 329}, 414(2004).

\bibitem{Rungta03} P. Rungta and C.M. Caves, Phys. Rev. A \textbf{67},
012307(2003).

\bibitem{Chen-Albeverio-Fei} K. Chen, S. Albeverio, and S. M. Fei,
Phys. Rev. Lett. \textbf{95}, 040504(2005).

\bibitem{Chen-Albeverio-Fei1} K. Chen, S. Albeverio, and S. M. Fei,
Phys. Rev. Lett. \textbf{95}, 210501(2005).

\bibitem{breuer}
Heinz-Peter Breuer, J. Phys. A: Math. Gen. \textbf{39}, 11847(2006)

\bibitem{Rains}E.M. Rains, Phys. Rev. A {\bf 60}, 179(1999).

\bibitem{Khasin06}M. Khasin, R. Kosloff, Phys. Rev. A \textbf{76}, 012304(2007).

\bibitem{E.M.Rains} E.M. Rains,   Phys. Rev. A  \textbf{60},  1(1999) .

\bibitem{S. Virmani} S. Virmani, M.F. Sacchi, M.B. Plenio, D. Markham,  Phys. Lett. A  {\bf 288}, 62(2001).

\bibitem{Yi} Y.X. Chen and D. Yang, Quantum Inf. Proc. \textbf{1}, 389(2002).

\bibitem{M.Khasin} M. Khasin, R. Kosloff and D. Steinitz, Phys. Rev. A \textbf{75}, 052325(2007).

\bibitem{FENG} F. Pan and G.Y. Lu, Int. J. Mod. Phys. B \textbf{20}, 1333(2006).

\bibitem{L. Lamata} L. Lamata, J. Le$\acute{\rm o}$n, D. Salgado and E. Solano, Phys. Rev. A \textbf{74}, 052336(2006).

\bibitem{Arun} A.K. Pati, Phys. Lett. A \textbf{278}, 118(2000).

\bibitem{B. M. Terhal} B. M. Terhal, Phys. Lett. A  {\bf 271}, 319(2000).

\bibitem{Philipp} P. Hyllus, O. G$\ddot{\rm u}$hne, D. Bru${\ss}$ and M. Lewenstein,
 Phys. Rev. A  {\bf 72}, 012321(2005).

\bibitem{Julio} J.I. de Vicente,  Quantum Inf. Comput. \textbf{7}, 624(2007).

\end{thebibliography}
\end{document}